\newcommand{\pT}{$p_{\mathrm{T}}$}
\newcommand{\ET}{$E_{\mathrm{T}}$}
\newcommand{\kT}{$k_{\mathrm{T}}$}
\newcommand{\dpT}{$\delta{p}_{\mathrm{T}}$}
\newcommand{\rhochem}{${\rho}_{\mathrm{ch+em}}$}

\newcommand{\RAA}{${R}_{AA}$}
\newcommand{\TAA}{${T}_{AA}$}

\documentclass[a4paper]{jpconf}
\usepackage{graphicx}
\begin{document}
\title{Full Jet Reconstruction in 2.76 TeV pp and Pb-Pb collisions in the ALICE experiment}

\author{Rosi Reed}

\address{Yale University, New Haven, CT 06520}

\ead{Rosi.Reed@Yale.edu}

\begin{abstract}
Measurements of the suppression of high-\pT~particles and the away-side jets from heavy-ion collisions at RHIC have shown that medium-induced energy loss affects partons produced in the early stage of a heavy-ion collision.  At LHC energies the initial production cross-section is much higher, which allows jets to be reconstructed with a wide kinematic range.  Measuring fully reconstructed jets by taking advantage of the ALICE Electromagnetic Calorimeter allows for a more differential investigation of the parton energy loss.   Parton energy loss will allow us to access key observables of the hot, dense matter created in heavy ion collisions.  The data presented was collected during the 2.76 TeV Pb-Pb runs, as well as baseline measurements from the 2.76 TeV pp run.   The procedures used to reconstruct jets and extract them from a fluctuating background will be discussed.  The procedure for quantifying the background with a limited acceptance will also be discussed.  These results are compared to pp measurements and simulations.
\end{abstract}

\section{Introduction}
One of the aims of studying heavy-ion collisions is to investigate how partons lose energy in the quark gluon plasma (QGP) that is formed in these collisions.  Hard scattered partons will fragment into a spray of hadrons, which is a jet.    The cross-section and fragmentation of jets has been extensively studied in proton-proton collisions so that  these measurements can be a reference for heavy-ion collisions.  In a heavy-ion collision, hard-scattered partons are produced early in the collision, which makes them an ideal probe of the QGP and for studying energy loss within the medium.  This parton energy loss will be reflected in the measured heavy-ion jet spectrum so that it is suppressed relative to a pp reference, which we call jet quenching  \cite{Wiedemann:2009sh}.  High-\pT~particle measurements, a proxy for jets, as well as other jet measurements have shown that the jet spectrum in heavy-ion collisions does deviate from what would be expected if the heavy-ion collision could be treated as a superposition of independent pp collisions \cite{Adams:2003kv, Adcox:2002pe, CMS:2012aa, Adler:2002tq,  Aamodt:2010jd, Adare:2006nr, AtlasJetQuench}.

With the added kinematic range at LHC energies and with improvements in jet finding techniques, it is now possible to fully reconstruct jets in heavy-ion collisions, including the charged component and the neutral components from ${\pi}^{0}$s, which should correspond better to the kinematics of the scattered parton than either charged jets or single particle measurements can.  The results reported in these proceedings are from data collected by the ALICE experiment in 2011 at an energy of $2.76$ TeV per nucleon pair.  Charged particles were reconstructed with the Time Projection Chamber (TPC) and the Inner Tracking System (ITS), while the neutral energy component from ${\pi}^{0}$s was reconstructed with the Electromagnetic Calorimeter (EMCal).  The Pb-Pb results are from the top 10\% most central events where the modification of the jet spectrum due to the QGP is at a maximum.

\section{Jets at ALICE}
The cells of the EMCal were clustered prior to inclusion in the jet finder by an algorithm that requires that each cluster only have a single local maximum.  Since charged particles, such as electrons or charged pions, will leave an energy signature in the EMCal the clusters were corrected for the hadronic contamination in order to prevent double counting of the charged energy.  All tracks with \pT$ > 0.15$ GeV/$c$ were propagated to the EMCal surface and then matched to clusters with \ET$ > 0.15$ GeV in the window $|\Delta\eta| < $0.015 and $|\Delta\varphi| <$ 0.03.  Each track was only matched to at most one cluster, though each cluster could have multiple tracks matched to it.  Then 100\% of the sum of the momenta of all matched tracks was removed from the cluster energy and then those clusters with \ET$ < 0.30$ GeV after hadronic correction were removed from the sample \cite{RReed1}.  

The collection of tracks and corrected EMCal clusters were then assembled into jets using the anti-\kT~or the \kT~algorithms \cite{JetFinder1, JetFinder2, JetFinder3, JetFinder4} with a resolution parameter of $R$ = 0.2.  Only those jets that were at least $R$ away from the EMCal boundaries of $|\eta| <$ 0.7 and 1.4 $<\phi < \pi$ and thus fully contained within the EMCal acceptance were used in the analyses described here.  The jets found by the anti-\kT~algorithm were used to determine the signal jets in both the pp and Pb-Pb analyses, the jets found by the \kT~algorithm were used to quantify the underlying event density in the Pb-Pb analysis.

\section{pp}
\label{sec:pp}

In order to fully quantify the modification of the jet spectrum in heavy-ion collisions due to the presence of the hot and dense QGP medium, the differential cross-section of fully reconstructed jets in pp collisions at $2.76$ TeV was measured.  Two different triggers were used to measure this spectrum, a minimum bias (MB) trigger and a EMCal trigger that required a single shower with ${E}_{\mathrm{T}}>$ 3 GeV.  The combination allowed the pp jet spectrum to be measured over a momentum range of 20 to 125 GeV/$c$.  The details of this trigger scheme can be found in \cite{RMa}.

\begin{figure}[htbp]
\begin{center}
 \includegraphics[width=0.5\textwidth]{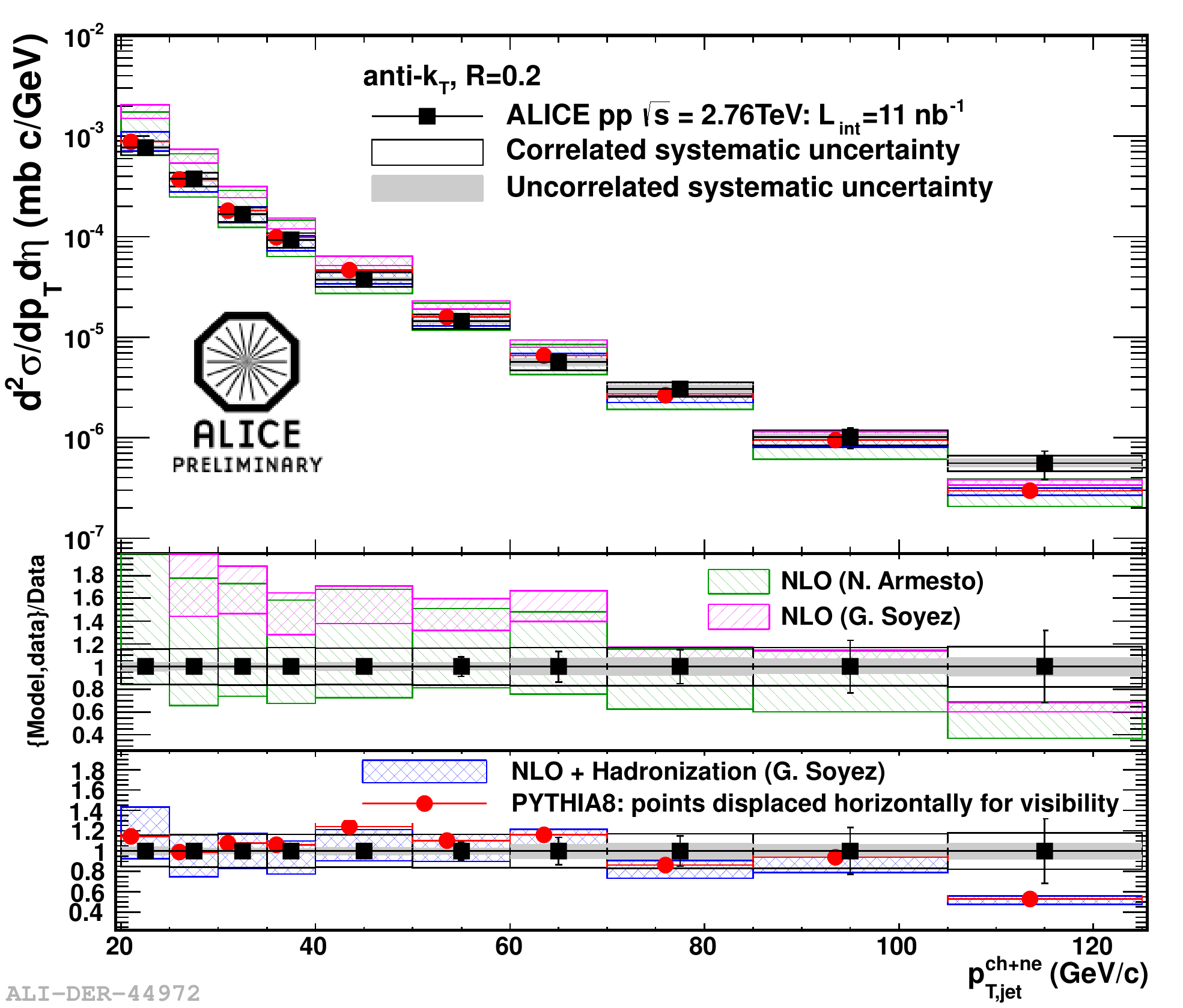}
 \end{center}
\caption{The solid boxes indicate the full inclusive jet differential cross section obtained with resolution parameter $R$ = 0.2 measured in pp collisions at 2.76 TeV data \cite{RMa}.  The solid circles represent a PYTHIA8 prediction.  The solid boxes are the uncorrelated systematic uncertainties on the cross-section, the open boxes are the correlated systematic uncertainties.  The hashed boxes indicate the NLO calculations (color online).}
\label{fig:ppPlot}
\end{figure}

The measured jets were corrected back to the particle level utilizing a simulation based bin-by-bin technique.  This was done by comparing the particle level jet spectra from a PYTHIA simulation to that the detector level spectrum, where the detector corrections were applied by use of GEANT.  The contributions to the correction can be divided into two categories: effects that smear the jet energy and effects that shift the jet energy. The effects that shift the jet energy are the hadronic correction, the tracking efficiency and unmeasured neutral particles.  The effects that smear the jet energy are the event-by-event fluctuations of the jet energy shift and the resolution of the detectors in measuring the jet constituents.  In this analysis, the jet energy shift was approximately 20\% with an uncertainty less than 4\% \cite{RMa}.  Figure \ref{fig:ppPlot} shows the inclusive differential jet cross section obtained with $R$ = 0.2 compared to a pQCD calculation at NLO and a PYTHIA8 prediction \cite{RMa}.  The calculations and simulations agree well with the measured spectrum which indicate that the process of jet formation is well understood in collisions where a QGP is not formed.
\section{Pb-Pb}

One of the main experimental challenges in heavy-ion collisions is removing the contribution from the underlying event from the jet spectrum.  This is done by determining the average background energy density, \rhochem, by finding the median of ${p}_{T,jet}/{A}_{jet}$, where ${A}_{jet}$ is the area and ${p}_{T,jet}$ is the uncorrected momentum of the \kT~jets.  This background was subtracted from the reconstructed momentum of the anti-\kT~ jets using the formula ${p}_{\mathrm{T},jet} = {p}_{\mathrm{T}}^{rec} - \rho\cdot{A}_{jet}$ \cite{JetFinder3, JetFinder4}.  However this does not account for the effect of the point-to-point background fluctuations which smear the jet energy.  These fluctuations are quantified by  \dpT~$= {p}_{\mathrm{T}}^{rec} - \rho\cdot{A}_{jet} - {p}_{\mathrm{T}}^{probe}$ \cite{ALICEbackground}.  The width of the distribution was determined to be 6.1 GeV/$c$ using a random cone method, and the effect of the smearing of the jet energy due to these background fluctuations is corrected through unfolding \cite{RReed1, SAiola1}.

\begin{figure}[htbp]
\begin{center}
\mbox{\includegraphics[width=0.45\textwidth]{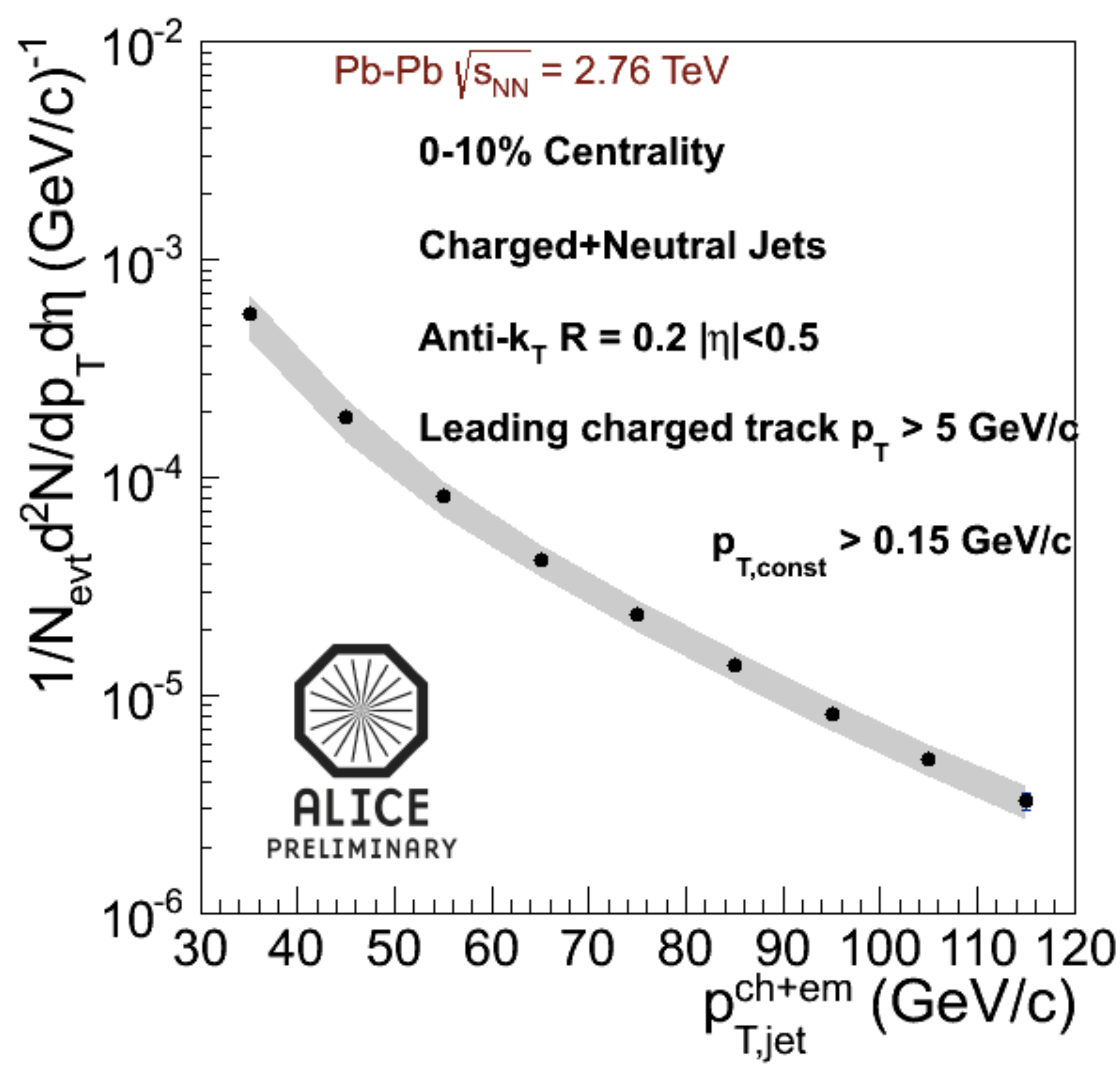}}
 \mbox{\includegraphics[width=0.45\textwidth]{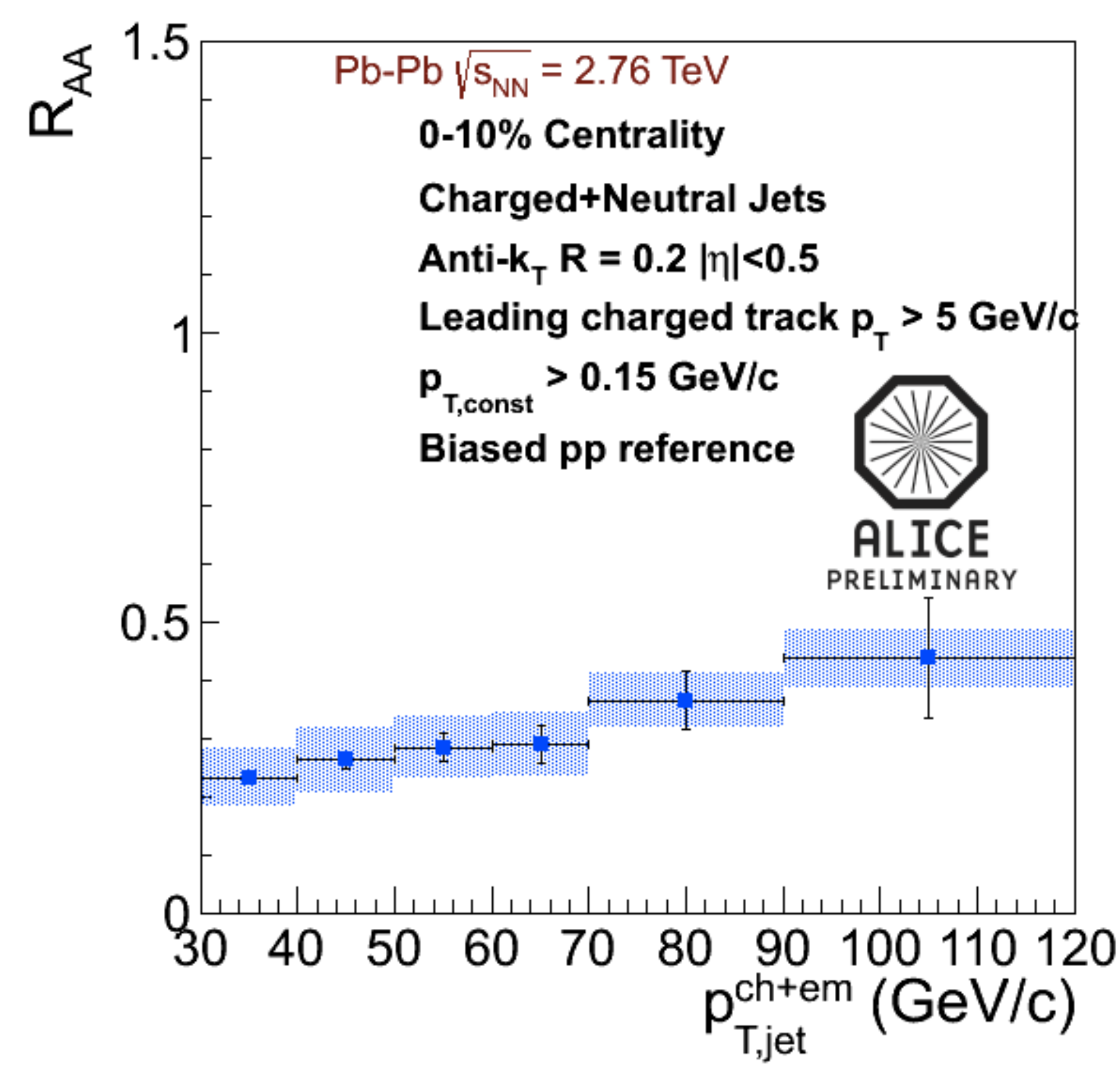}}
 \end{center}
 \caption{On the left is the fully reconstructed jet spectrum from 0-10\% central events in Pb-Pb collisions measured by ALICE at $\sqrt{{S}_{NN}}$ = 2.76 TeV.  The jets were constrained to the ALICE EMCal and biased so that their leading track had \pT$ >$ 5 GeV/$c$.  The band around the data points represents the systematic uncertainty. The \RAA~for $R$ = 0.2 biased jets in 0-10\% central events is on the right.  The statistical and systematic uncertainties from the pp and Pb-Pb analyses are added in quadrature.  The combined systematics are shown as dotted boxes. } 
\label{fig:RAA}
\label{fig:spectra}
\end{figure}

Unfolding will also correct the measured background subtracted jet spectrum for detector effects as outlined in Section \ref{sec:pp}.  In order to have a stable result it was necessary to bias the collection of measured jets by requiring that the jets have a track with \pT$ >$ 5 GeV/$c$.  This reduces the contribution to the measured spectrum due to the combinatorial background.  The resulting spectrum is shown in Figure \ref{fig:spectra}.  The systematic uncertainty is \pT~dependent with a value of 18\% at 60 GeV/$c$, and dominated by the hadronic correction, the tracking efficiency and the unfolding process itself.  Also see \cite{SAiola1} for more details.

\section{\RAA}

The nuclear modification factor, \RAA, is defined as the ratio of the jet spectrum from Pb-Pb collisions as shown in Figure \ref{fig:spectra} to the differential cross-section from pp scaled by \TAA, where \TAA~is the ratio of the number of binary collisions over the total inelastic pp cross-section.  \RAA~is constructed so that it would be equal to one if there is no suppression or enhancement of a signal in heavy ion collisions as compared to pp collisions.  However since a bias was necessary to measure the jet spectrum in Pb-Pb collisions, if the unbiased reference spectra shown in  Figure \ref{fig:ppPlot} was used, \RAA~could be constructed as less than 1 even if there is no modification of the heavy ion jet spectrum due to nuclear effects.  In order to avoid this, a biased pp reference spectrum was constructed by applying the same method outlined in \cite{RMa}.  The resulting \RAA~can be seen in Figure \ref{fig:RAA}.  The systematic and statistical uncertainties from the Pb-Pb and pp analysis were added in quadrature.  We have found that fully reconstructed jets are suppressed in the 10\% most central heavy-ion collisions in a \pT~dependent manner. This measurement is not inconsistent with conservation of energy as the energy lost by the jets could be recovered elsewhere, for instance at a larger $R$ or at a lower momentum. These results agree well the the previous charged only results \cite{MartaHP}, though the energy scale is not the same in both cases so a direct comparison is not possible.  



\section{Conclusions}
In this proceedings we have reported on a corrected fully reconstructed jet spectrum from the 2011 Pb-Pb data from the 10\% most central evens in ALICE.  The spectra combined with the differential cross section measured in pp collisions allowed us to determine the nuclear modification factor.  We have seen that $R$ = 0.2 jets are suppressed in these most central events and that this suppression has a \pT~dependence.  This result is still consistent with conservation of energy laws as the lost energy may be recoverable at larger angles or lower momentum than what are measuring in this analysis.  Additional measurements, such as determining the dependence of the nuclear modification factor with centrality, event plane and $R$, will be necessary to fully quantify the measured jet quenching.  These measurement can then be compared to available energy loss models.  These results are consistent with the CMS $R$ = 0.2 results from the 5\% most central events \cite{CMS}, though there are variations between the analyses such as the background determination.  Jet analyses in heavy-ion collisions are experimentally difficult, in particular the background subtraction and correction of fluctuations can be performed in different ways.  Therefore it is important for all of the LHC heavy-ion experiments to converge upon a single consistent jet measurement.

\end{document}